\documentclass[aps,prd,preprint,superscriptaddress,showpacs]{revtex4}
\usepackage{graphics}
\usepackage{epsfig}
\usepackage{latexsym}
\usepackage{colordvi}
\usepackage{amsmath}
\usepackage{amssymb}
\def\bi{\bibitem}
\def\la{\langle}\def\ra{\rangle}
\def\bi{\bibitem}
\def\lsim{\mathrel{\rlap{\lower3pt\hbox{\hskip1pt$\sim$}}
     \raise1pt\hbox{$<$}}} %less than or approx. symbol
\def\gsim{\mathrel{\rlap{\lower3pt\hbox{\hskip1pt$\sim$}}
     \raise1pt\hbox{$>$}}}

\def\be{\begin{eqnarray}}\def\ba{\begin{eqnarray}}
\def\ee{\end{eqnarray}}\def\ea{\end{eqnarray}}
\def\ben{\begin{enumerate}}\def\bitem{\begin{itemize}}
\def\een{\end{enumerate}}\def\eitem{\end{itemize}}

\def\VMD0{VMD$^{gs}$}\def\VMDinf{VMD$^\infty$}

\begin{document}
\preprint{\parbox[b]{1in}{ \hbox{\tt PNUTP-07/A06} {\tt KIAS-P07071} {\tt IC/2007/121} }}

\title{Nucleon Form Factors and Hidden Symmetry\\ in Holographic QCD}

\author{Deog Ki Hong}
\email[E-mail: ]{dkhong@pusan.ac.kr} \affiliation{Department of
Physics, Pusan National University,
             Busan 609-735, Korea}

\author{Mannque Rho}
\email[E-mail: ]{ Mannque.Rho@cea.fr} \affiliation{Service
de Physique Th\'eorique, CEA Saclay, 91191
Gif-sur-Yvette, France}

\author{Ho-Ung Yee}
\email[E-mail: ]{hyee@ictp.it} %\\
\affiliation{School of Physics,
Korea Institute for Advanced Study, Seoul 130-722, Korea}
\affiliation{The Abdus Salam International Centre for
Theoretical Physics, Strada Costiera 11, 34014, Trieste, Italy} \vspace{0.1in}

\author{Piljin Yi}
\email[E-mail: ]{piljin@kias.re.kr} \affiliation{School of Physics,
Korea Institute for Advanced Study, Seoul 130-722, Korea}

\vspace{0.1in}

\date{\today}

\begin{abstract}

The vector dominance of the electromagnetic  form factors both for
mesons and baryons arises naturally in holographic QCD, where both
the number of colors and the
't Hooft coupling are taken to be very large,
offering a bona-fide derivation of the notion of vector dominance.
The crucial ingredient for this is the infinite tower of vector mesons in the
approximations made which share features that are characteristic of the quenched
approximation in lattice QCD. We approximate the infinite sum by contributions
from the lowest four vector
mesons of the tower which turn out to saturate the charge
and magnetic moment sum rules within a few percent and compute them
totally free of unknown parameters for momentum transfers $Q^2\lsim
1$ GeV$^2$. We identify certain observables that can be reliably computed
within the approximations and others that are not, and discuss how the
improvement of the latter can enable one to bring holographic QCD closer to QCD proper.

\end{abstract}
%%% 12.38.-t Quantum chromodynamics
%%% 11.25.Tq Gauge/string duality
%%% 12.38.Aw General properties of QCD
%%% 14.20.Dh protons and neutrons
%%% 11.10.Kk Field theory in dimension other than four

\maketitle
\newpage
\section{Introduction}
Recently, there has been active study on the gravity dual models of quantum
chromodynamics (QCD), named as holographic QCD (hQCD).
One of the consequences of hQCD is the vector meson
dominance in the low energy dynamics of QCD, where not just the
ground-state vector mesons $V^{gs}=\rho, \omega, \phi$ contribute,
as originally proposed by Sakurai in the 1960s~\cite{Sakurai:1969ss},
but the whole tower of vector mesons (denoted $V^\infty$), including
all excited states, do contribute. If this new vector meson
dominance is verified experimentally, it will indicate strongly that
QCD has a hidden symmetry, which is best described by Yang-Mills
gauge fields in a five-dimensional spacetime with a warped factor
dependent on the extra fifth dimension, and which, reduced to 4
dimensions, manifest themselves as an infinite tower of vector and
axial vector mesons. How the 5D spacetime is curved at the hadronic scale
can be inferred
directly from the spectra of vector mesons and their role in the
response functions of hadrons to external fields.

The vector meson dominance with the ground-state mesons ($V^{gs}$)
has been a powerful tool in studying electroweak structure of
hadrons in nuclear and hadron physics. It has met with a remarkable
phenomenological success, ranging from electromagnetic form factors
of light-quark hadrons to deep-inelastic scattering in the
diffraction region of small $x=Q^2/W^2\ll 1$. (See \cite{VD-rev} for
a recent discussion on this.) On the theoretical side, however, the
situation has been less than satisfactory. First of all there is no
bona-fide ``derivation" of VMD$^{gs}$ with the ground-state vector
mesons from first principles. The formulation of \VMD0 employing an
operator identity known as current-field
identity~\cite{GZ1961,KLZ1967} -- which was anchored on gauge
principle -- gave a natural explanation of ``universality" of
electromagnetic coupling $g_{\rho\pi\pi}=g_{\rho\rho\rho}=g_{\rho
NN}$. However there was a glaring defect in describing nucleon
electromagnetic (EM) form factors in terms of \VMD0, which destroyed
the notion of universality. This shortcoming has become more
prominent with the advent of precision measurements of nucleon form
factors at the JLab.

The problem,  briefly stated, is as follows. While \VMD0 works
remarkably well to large momentum transfers for the pion form factor
in both the space-like and time-like regimes, it fails badly for the
nucleon. Although the early nucleon form factor measurements were
interpreted as empirical evidence for an isoscalar vector meson,
$\omega\rightarrow 3\pi$, by Nambu~\cite{nambu} in 1957, and for an
isovector meson, $\rho^0\rightarrow 2\pi$, by Frazer and
Fulco~\cite{frazer-fulco} in 1959, it has subsequently been shown
that nucleon EM form factors cannot be described satisfactorily by
the exchange of $V^{gs}$ alone but requires an important additional
component representing an ``intrinsic core" of size $\sim 0.3-0.4$
fm~\cite{orear,petronzio}. A two-component picture implementing the
latter to the standard \VMD0 has been modeled by including a direct
photon coupling to the nucleon that has an intrinsic core of an
unspecified source~\cite{IJL} or a bag of confined chiral
quarks~\cite{chiralbag}. The consequence of this two-component
picture was that the universality encoded in \VMD0 is lost, i.e.,
$g_{\rho\pi\pi}\neq g_{\rho NN}$ and hence vector dominance is
violated.

It was shown in \cite{Hong:2007ay} that in the holographic, instanton picture of
the nucleon, which is effectively the Skyrmion picture drastically
modified by the inclusion of infinite number of vector mesons and
axial-vector mesons,
% in the skyrmion description of
%the nucleon embedded in an infinite tower of vector mesons that
%results from an instanton structure in 5D in the Sakai-Sugimoto (SS)
a full vector dominance re-emerges. This  restores the
universality relation -- lost with $V^{gs}$ alone -- in a form that
involves the whole tower. In this paper, we report how the model for
the nucleon EM form factors fares with nature at low momentum
transfers. What we are doing here can be considered as a
$prediction$ -- and not a postdiction -- of the hQCD model since the
calculation involves {\it no free parameters}: all the pertinent
parameters of the action in the approximations adopted, i.e., large
$N_c$ and $\lambda=g_s^2 N_c$ limits, are completely fixed in the
meson sector \cite{Sakai:2004cn}. This model was found
\cite{Hong:2007ay} to give a satisfactory description of chiral
dynamics that can be reliably treated -- such as the axial coupling
constant, isovector magnetic moment etc. -- in the quenched
approximation of QCD.

\section{Vector Dominance for Nucleon Form Factors}

The nucleon form factors are defined from the matrix elements of the
external currents,
\begin{equation}
\left<p^{\prime}\right|J^{\mu}(x)\left|p\right>=e^{iqx}\,\bar
u(p^{\prime})\,{\cal O}^{\mu}(p,p^{\prime})\,u(p) \:,
\end{equation}
where $q=p^{\prime}-p$ and $u(p)$ is the nucleon spinor of momentum $p$.
By the Lorentz invariance and the current
conservation we expand the operator ${\cal O}^{\mu}$, assuming the CP invariance, as
\begin{equation}
{\cal
O}^{\mu}(p,p^{\prime})=\gamma^{\mu}\left[\frac12F_1^s(Q^2)+F^a_1(Q^2)\tau^a\right]
+i\frac{\sigma^{\mu\nu}}{2m_N}q_{\nu}
\left[\frac12F_2^s(Q^2)+F^a_2(Q^2)\tau^a\right]\:,
\end{equation}
where $m_N\simeq 940~{\rm MeV}$ is the nucleon mass and $\tau^a=\sigma^a/2$.
$F_1^s$ and $F_2^s$ are the Dirac and Pauli form factors
for iso-scalar current respectively, and  $F_1^a$, $F_2^a$ are for iso-vector currents.

Being matrix elements, the form factors contain all one-particle irreducible diagrams
for two nucleons and one external current %, given as
%\begin{equation}
%\left<p^{\prime}\right|J^{\mu}(x)\left|p\right>=\left<p^{\prime}\right|
%\frac{\delta}{\delta A_{\mu}} e^{iS_{\rm eff}[A]}\left|p\right>,
%\end{equation}
%where $W[J]$ is the generating functional for the current $J_{\mu}$.
and thus very difficult to calculate from QCD. However, the anti-de
Sitter/conformal field theory (AdS/CFT) correspondence, or gravity/gauge theory correspondence, found in
certain types of string theory, enables us to compute such
non-perturbative quantities like hadron form factors with mild
approximations~\cite{Hong:2004sa,Brodsky:2007hb}.

According to this correspondence, the low energy effective action of the gravity
dual of QCD becomes the generating functional for the correlators
of an operator ${\cal O}$ in QCD in the large $N_c$ limit:
\begin{equation}
e^{iS_{5D}^{\rm eff}[\phi(\epsilon,x)]}=\left<\exp\left[i\int_x
\phi_0\,{\cal O}\right]\right>_{\!\!\rm QCD}\,,\label{correspondence}
\end{equation}
where $\phi(z,x)$ is a bulk field, acting as a source for ${\cal O}$
when evaluated at the UV boundary $z=\epsilon$.
Furthermore the normalizable modes of the bulk field are identified as the physical states
in QCD, created by the operator ${\cal O}$.

A gravity dual of low energy QCD  with  massless flavors
is proposed in the quenched approximation by Sakai and Sugimoto (SS)~\cite{Sakai:2004cn}.
Later, the holographic dual model of spin-$1\over 2$ baryons, or nucleons, in
the SS model with two flavors is constructed~\cite{HRYY-PR,Hong:2007ay}
by introducing a bulk baryon field, whose effective
action is given in the conformal coordinate $(x,w)$ as
\begin{eqnarray}
S_{5D}^{\rm eff}=\int_{x,w}\left[-i\bar{\cal B}\gamma^m D_m {\cal B}
-i m_b(w)\bar{\cal B}{\cal B} +\kappa(w)
\bar{\cal B}\gamma^{mn}F_{mn}^{SU(2)_I}{\cal B}+\cdots \right]+S_{\rm meson},
\label{5dfermion1}
\end{eqnarray}
where ${\cal B}$ is the 5D bulk baryon field, $D_m$ is the gauge covariant derivative
and $S_{\rm meson}$ is the effective action for the mesons~\cite{foot3}.
Using the instanton nature of baryon, the coefficients $m_b(w)$ and $\kappa(0)$
can be reliably calculated in string Theory~\cite{HRYY-PR,Hong:2007ay}.
Especially, the coefficient of the magnetic coupling is
estimated to be
\begin{equation}
\kappa(w)\simeq \frac{0.18N_c}{M_{KK}}\,,
\end{equation}
where $M_{KK}\simeq 0.94$ GeV is the ultraviolet cutoff of the SS model.
The ellipsis in the effective action (\ref{5dfermion1}) denotes  higher derivative operators,
whose coefficients are difficult
to estimate, but are suppressed
at low energy, $E<M_{KK}$.
The important point is that the magnetic coupling involves only the non-Abelian part of flavor
symmetry $SU(2)_I$, with Abelian $U(1)_B$ being absent, due to the non-Abelian nature of instanton-baryons.

Though the exact correspondence between the gravity dual and QCD is not established yet,
we compute the electromagnetic (EM) form factors for the nucleons
in the SS model by assuming  the correspondence.
We first need to identify the dual bulk field of the external EM current, which is
nothing but the bulk photon field.
Since the electric charge operator is the sum of isospin and the baryon operator,
\begin{equation}
Q_{\rm em}=I_3+\frac12 B,
\end{equation}
we have to identify a combination of $A_{\mu}^3$ and $A_{\mu}^B$, the third component of the isospin
gauge field and the $U(1)_B$ gauge field, respectively, as the photon field.
Then all baryon bilinear operators in the effective action
that couple to either $U(1)_B$ gauge fields or $SU(2)_I$ gauge
fields will contribute to the EM form factors.
%For a constant source $J_{\mu}(x)=e^{iqx}$

We now write
the (nonnormalizable) photon field as
\begin{equation}
A_{\mu}(x,w)=\int_q\,A_{\mu}(q)A(q,w)\,e^{iqx}\:,
\end{equation}
with boundary conditions
that $A(q,w)=1$ and $\partial_wA(q,w)=0$ at the UV boundary, $w=\pm w_{\rm max}$ and
the (normalizable) bulk baryon field as
\begin{equation}
{\cal B}(w,x)=\int_p\left[f_L(w)u_L(p)+f_R(w)u_R(p)\right]e^{ipx}\,.
\end{equation}
These 5D wave functions, $A(q,w)$ and $f_{L,R}(w)$, are determined by solving the equation of motion from our action (\ref{5dfermion1}).
Then, using the correspondence (\ref{correspondence}), one can read off
the form factors.
We find for nucleons
the Dirac form factor $F_1(Q^2)=F_{1{\rm min}}\,Q_{\rm em}+F_{1{\rm mag}}\,I_3$ with ($Q^2\equiv -q^2$)
\begin{eqnarray}
F_{1{\rm min}}(Q^2)&=&\int_{-w_{max}}^{w_{max}} dw\,\left|f_L(w)\right|^2\,A(q,w)\:,\label{ff_1}\\
F_{1\rm mag}(Q^2)&=&2\int_{-w_{max}}^{w_{max}}
dw \kappa(w)\left|f_L(w)\right|^2\partial_w A(q,w)\:,\label{ff_1m}
\end{eqnarray}
where $F_{1{\rm min}}$ is from the minimal coupling, and $F_{1\rm mag}$ the magnetic coupling.
Similarly the Pauli form factor is given as $F_2(Q^2)=F^3_{2}(Q^2)\,I_3$ with
\begin{eqnarray}
F^3_{2}(Q^2)&=&4\,m_N\,
\int_{-w_{max}}^{w_{max}} dw\,
\kappa(w)f_L^*(w)f_R(w)\,A(q,w)\:,
\label{ff_2}\end{eqnarray}
which comes solely from the magnetic coupling.
We note that the form factors (\ref{ff_1}), (\ref{ff_1m}) and (\ref{ff_2}) have corrections
coming from the higher order operators in the effective action~(\ref{5dfermion1}),
which are suppressed by powers of  $E/M_{KK}$ at low energy. However we emphasize that
our result contains full quantum effects in the large $N_c$ limit,
because it is computed from the generating functional for one-particle
irreducible correlation functions, summed over all loops, prescribed by the AdS/CFT correspondence.

Alternatively, we can replace the form factors by
an infinite sum of vector-meson
exchanges~\cite{Hong:2004sa,footnote00}, if we expand the nonnormalizable  photon field
in terms of the normalizable vector meson $\psi_{2k+1}$ of mass $m_{2k+1}$ as
\begin{equation}
A(q,w)=\sum_k\frac{g_{v^{(k)}}\psi_{(2k+1)}(w)}{Q^2+m_{2k+1}^2}\:,
\label{expansion}
\end{equation}
where the decay
constant of the $k$-th vector mesons  is given as
$g_{v^{(k)}}=m_{2k+1}^2\zeta_k$ with
\begin{equation}
\zeta_k=\frac{\lambda N_c}{108\pi^3}M_{KK}\int_{-w_{max}}^{w_{max}}
 dw\frac{U(w)}{U_{KK}}\,\psi_{(2k+1)}(w) \:,
\end{equation}
where $U_{KK}$ is a parameter of the SS model and
\begin{equation}
dw=\frac32\frac{U_{KK}^{1/2}}{M_{KK}}\frac{dU}{\sqrt{U^3-U_{KK}^3}}\,.
\end{equation}
The resulting EM form factors then take the
form
%with $g_1^{(k)}=g^{(k)}_{V,min}+\frac12g_{V,mag}^{(k)}$
\begin{eqnarray}
F_{1}(Q^2)&=&F_{1{\rm min}}\,Q_{\rm em}+F_{1{\rm mag}}\,\tau^3=
\sum_{k=1}^{\infty} \left(g^{(k)}_{V,min}Q_{\rm em}
+g_{V,mag}^{(k)}\,\tau^3\right)\frac{\zeta_k m_{2k+1}^2}
{Q^2+m_{2k+1}^2}\,,\\
F_2(Q^2) &=&F_2^3(Q^2)\,\tau^3= \,\tau^3\sum_{k=1}^{\infty}
{g_2^{(k)}\zeta_k m_{2k+1}^2\over Q^2+m_{2k+1}^2}\,,
\end{eqnarray}
where
\begin{eqnarray}
g_{V,min}^{(k)}&=&\int_{-w_{max}}^{w_{max}} dw\,\left|f_L(w)\right|^2
\psi_{(2k+1)}(w)\\
g_{V,mag}^{(k)}&=& 2\int_{-w_{max}}^{w_{max}}
 dw \,\kappa(w)\left|f_L(w)\right|^2
 \partial_w \psi_{(2k+1)}(w)\:,\nonumber\\
 g_2^{(k)}&=&
 4m_N \int_{-w_{max}}^{w_{max}}
dw \,\kappa(w) f_L^*(w)f_R(w) \psi_{(2k+1)}(w)\:.
\end{eqnarray}

The authors of~\cite{Hong:2007ay} noted that the sum rules for the
electric charge and the magnetic moment given in the large $N_c$ and
$\lambda$ limit are well saturated by first four vector mesons.
After shifting $N_C\to N_C+2$ for the magnetic coupling as described
in \cite{Hong:2007ay}~\cite{footnote0}, one finds that the sum rules
are saturated for protons (and similarly for neutrons) within a few
\% when we take $N_C=3$:
\begin{eqnarray}
F^p_1(0)&\equiv&1\simeq \sum_{k=1}^4\left(g^{(k)}_{V,min}+{\frac12}\cdot{N_C+2 \over N_C}\cdot g_{V,mag}^{(k)}\right)
\zeta_k=1.04,\nonumber\\ F^p_2(0)
&\equiv&\mu_p-1\simeq{\frac12}\cdot{N_C+2 \over N_C}\cdot\sum_{k=1}^4g_2^{(k)}\zeta_k=1.66\,.
\end{eqnarray}
Recovering the mass unit $M_{KK}$, we obtain, taking $N_c=3$ and
$f_\pi\simeq 93$ MeV as determined in the meson sector, from Table~2
of~\cite{Hong:2007ay}
\begin{eqnarray}
\!\!F^p_1(Q^2)\!\!&\simeq&\!\! \frac{0.958~M_{KK}^2}{Q^2+0.67~M_{KK}^2}
-\frac{1.230~M_{KK}^2}{Q^2+2.87~M_{KK}^2}
-\frac{0.628~M_{KK}^2}{Q^2+6.59~M_{KK}^2}
+\frac{1.585~M_{KK}^2}{Q^2+11.8~M_{KK}^2}\label{4}\\
\!\!\!\!\!\!\!\!F_{1{\rm mag}}(Q^2)\!\!&\simeq&\!\! -\frac{0.248~M_{KK}^2}{Q^2+0.67~M_{KK}^2}
+\frac{2.602~M_{KK}^2}{Q^2+2.87~M_{KK}^2}
-\frac{5.777~M_{KK}^2}{Q^2+6.59~M_{KK}^2}
+\frac{5.153~M_{KK}^2}{Q^2+11.8~M_{KK}^2}\label{4_1}\\
\!\!F^p_2(Q^2)\!\!&\simeq&\!\!\frac{1.855~M_{KK}^2}{Q^2+0.67~M_{KK}^2}
-\frac{4.587~M_{KK}^2}{Q^2+2.87~M_{KK}^2}
+\frac{4.547~M_{KK}^2}{Q^2+6.59~M_{KK}^2}
-\frac{2.390~M_{KK}^2}{Q^2+11.8~M_{KK}^2}.\label{5}
\end{eqnarray}
Equations (\ref{4}), (\ref{4_1}) and (\ref{5})
constitute the main ingredients for the analysis that follows.

The observable quantities we are interested in are the Sachs form factors for
protons and neutrons defined
by
\begin{eqnarray}
G_M^p(Q^2)&=&F^p_{1}(Q^2)+F^p_2(Q^2)\:,\label{GM}\\
G_E^p(Q^2)&=&F^p_{1}(Q^2)-\frac{Q^2}{4m_N^2}F^p_2(Q^2)\:,\label{GE}\\
G_M^n(Q^2)&=&-{1\over 2}F_{1\rm mag}(Q^2)-F^p_2(Q^2)\:,\\
G_E^n(Q^2)&=&-{1\over 2}F_{1\rm
mag}(Q^2)+\frac{Q^2}{4m_N^2}F^p_2(Q^2),
\end{eqnarray}
where we used the fact that the Pauli form factor of neutron is $F_2^n(Q^2)=-F^p_2(Q^2)$
in holographic QCD, found by the authors~\cite{HRYY-PR,Hong:2007ay}.
\section{Numerical Results}
Before we make the estimates of various physical quantities,
let us briefly review the standard practice in comparing theoretical
(model) calculations with experiments.

 All the parameters in our approach are fixed in the meson sector,
 so there are no free parameters to adjust. In phenomenological models based on the vector dominance by the ground state $V^{gs}$~\cite{IJL} or on the skyrmion core surrounded by the $V^{gs}$ cloud~\cite{holzwarth}, however, one takes into account several features that are extraneous to the models. For instance, perturbative QCD tells us that at asymptotic momentum transfer, $F_i(Q^2)$ should scale as $F_1\sim Q^{-4}$ and $F_2\sim Q^{-6}$. Our form factors (\ref{4}) and (\ref{5}) do not possess these scaling features. Furthermore, even if one includes one or more excited vector mesons in the model, one needs to account at least for the fact that the ground-state $\rho$ has a large width. All these are included by hand in phenomenological model fits. If we wished to fit the experimental data to $Q^2\sim 10\ {\rm GeV}^2$ available in the literature~\cite{data}, these properties should also be taken into account. The hQCD we are studying cannot handle any of these at present since they require calculating $1/N_c$ and $1/\lambda$ corrections and include short-distance interactions given by perturbative QCD. Since our aim here is to see how the theory in the given approximations fares with no unknown parameters, we will eschew introducing arbitrary phenomenological factors. We will therefore limit our kinematics to $Q^2\lsim M_{KK}^2 \sim 1$ GeV$^2$.
\subsection{Charge and Magnetic Radii}
Since we have seen that static quantities involving $F_i (0)$ come
out in good agreement with experiments~\cite{Hong:2007ay}, the next
low-momentum quantity we can calculate are the electric charge
radius $r_{e}$ and magnetic radius $r_{m}$. We will focus on the
proton structure. Taking $M_{KK}\approx m_N\simeq 0.94$ MeV as fixed
in the meson sector~\cite{Sakai:2004cn}, we readily obtain \be
{r_{pe}^2} &\equiv& -6\,\left.\frac{{\rm d}}{{\rm
d}Q^2}\left[\ln\,G_E^p(Q^2)\right]\right\vert_{Q^2=0}\simeq
\left(0.796~{\rm fm}\right)^2,\label{Re}\\
{r_{pm}^2} &\equiv& -6\,\left.\frac{{\rm d}}{{\rm
d}Q^2}\left[\ln\,G_M^p(Q^2)\right]\right\vert_{Q^2=0}\, \simeq
\left(0.744~{\rm fm}\right)^2.\label{Rm}
\ee
These are to be compared with the experimental values~\cite{data}
 \be
{\tilde r}_{pe}^{exp}= 0.895 \pm 0.018~{\rm fm},\,\,\, {\tilde
r}_{pm}^{exp}= 0.855 \pm 0.035~{\rm fm}.
 \ee

We first note that both of the predicted radii are $\sim$ 0.1 fm smaller than the
experimental values. This deviation
is expected since our estimate is reliable only when the number of
colors and also the 't~Hooft coupling $\lambda$ are very large. In
the SS model of QCD, the ``core" radius of the nucleons scales as
$1/\sqrt{\lambda}$ and therefore the resulting core size $\Delta r \sim 0.1$
fm must be due to the subleading corrections, which are extremely
difficult to estimate. What is noteworthy is that the core size
which comes out to be $\sim 0.4$ fm when only the ground-state vector
mesons $V^{gs}$ are taken into account ~\cite{IJL}, shrinks to $\sim 0.1$ fm in the presence of the tower of
vector mesons, here encapsulated in the three higher-lying members.
The full account of the tower may shrink the ``core"
size further with the higher tower playing the role of a major part of the
intrinsic core or quark-bag degrees of freedom.

We should note however that the underestimate of the root-mean-square radii
while static quantities, e.g., magnetic moments, $g_A$ etc. come out close to experimental values~\cite{HRYY-PR} is a generic feature of quenched approximations as noticed in quenched lattice calculations of vector and axial nucleon form factors ~\cite{sasaki-yamazaki}.

\subsection{$\mu_p G^p_E (Q^2)/G^p_M(Q^2)$}
\begin{figure}[ht]
\centerline{\includegraphics[scale=0.6]{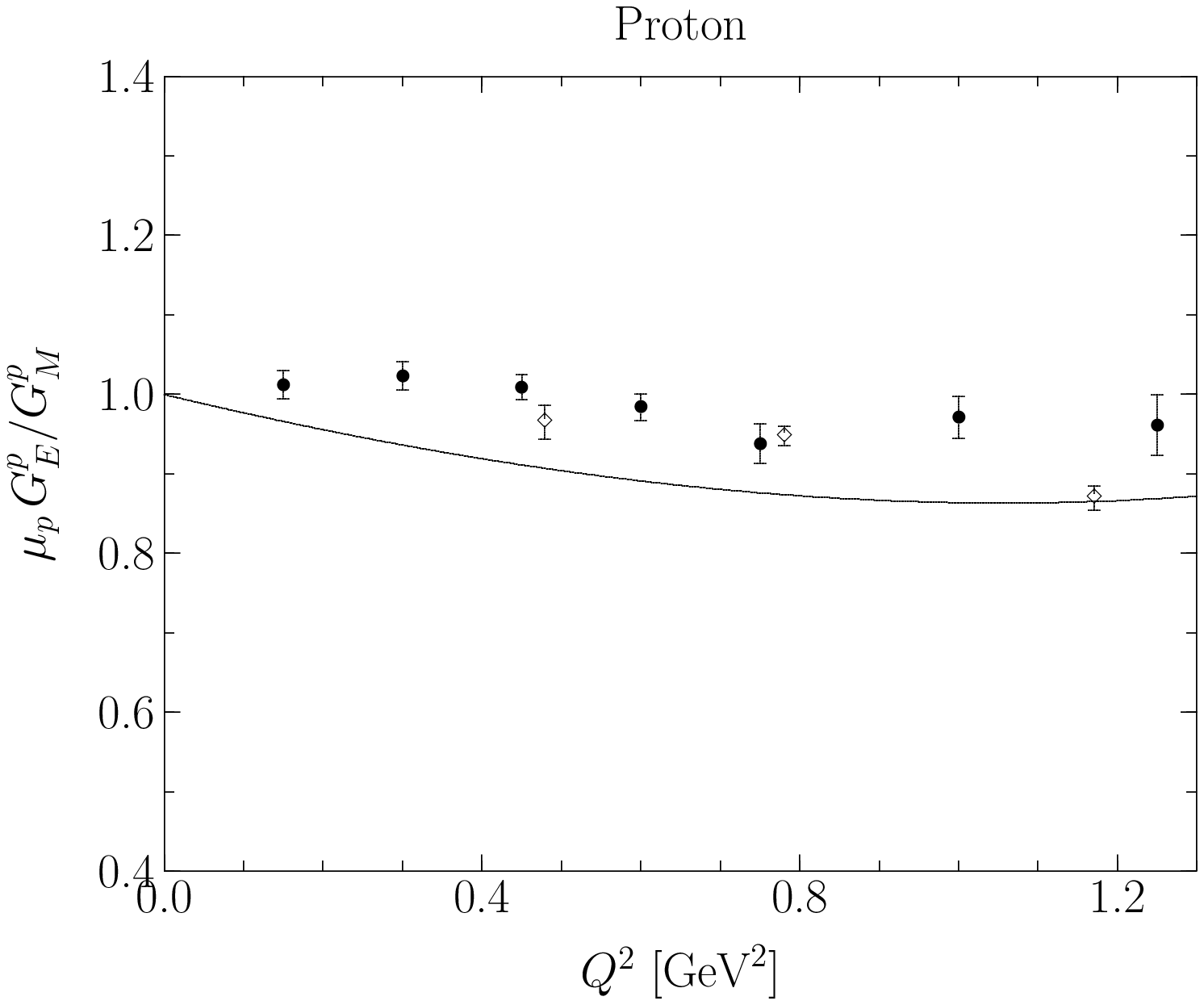}}%
\caption{The open cicles are the polarization measurements at
JLab~\cite{Jones:1999rz} and the filled circles are the data taken
from~\cite{Walker:1993vj}. The solid line is the prediction in the
SS model. }
 \label{fig2}
\end{figure}

We expect the ratios of form factors to be less sensitive than the form factors themselves to $1/N_c$ and $1/\lambda$ corrections and also to additional form factors representing asymptotic scaling which manifest themselves in the ``core size." We therefore look at the ratio
\be
R_1(Q^2)=\mu_p G^p_E (Q^2)/G^p_M(Q^2).
\ee
The result is plotted in Fig.\ref{fig2}.
Given that our calculations are valid in the large $N_c$ and
$\lambda$ limits and also in the chiral limit, and above all that there
are no free parameters~\cite{footnote1}, the agreement with
experiments within, say, $\lsim 10\%$ is quite surprising.
%and lends
%further support to the applicability of hQCD for chiral dynamics
%claimed in \cite{Hong:2007ay}.
However, the fact that the corrections cancel out largely in the ratio suggests that the $R_1$
belongs to the class of observables for which quenched approximations are applicable.
The $\lsim 10\%$ underestimate can be readily understood as explained below.
\subsection{$G^p_M (Q^2)/\mu_p G_D (Q^2)$ and $G^p_E (Q^2)/G_D (Q^2)$}
\begin{figure}[ht]
\vskip 0.1in \centerline{\includegraphics[scale=0.7]{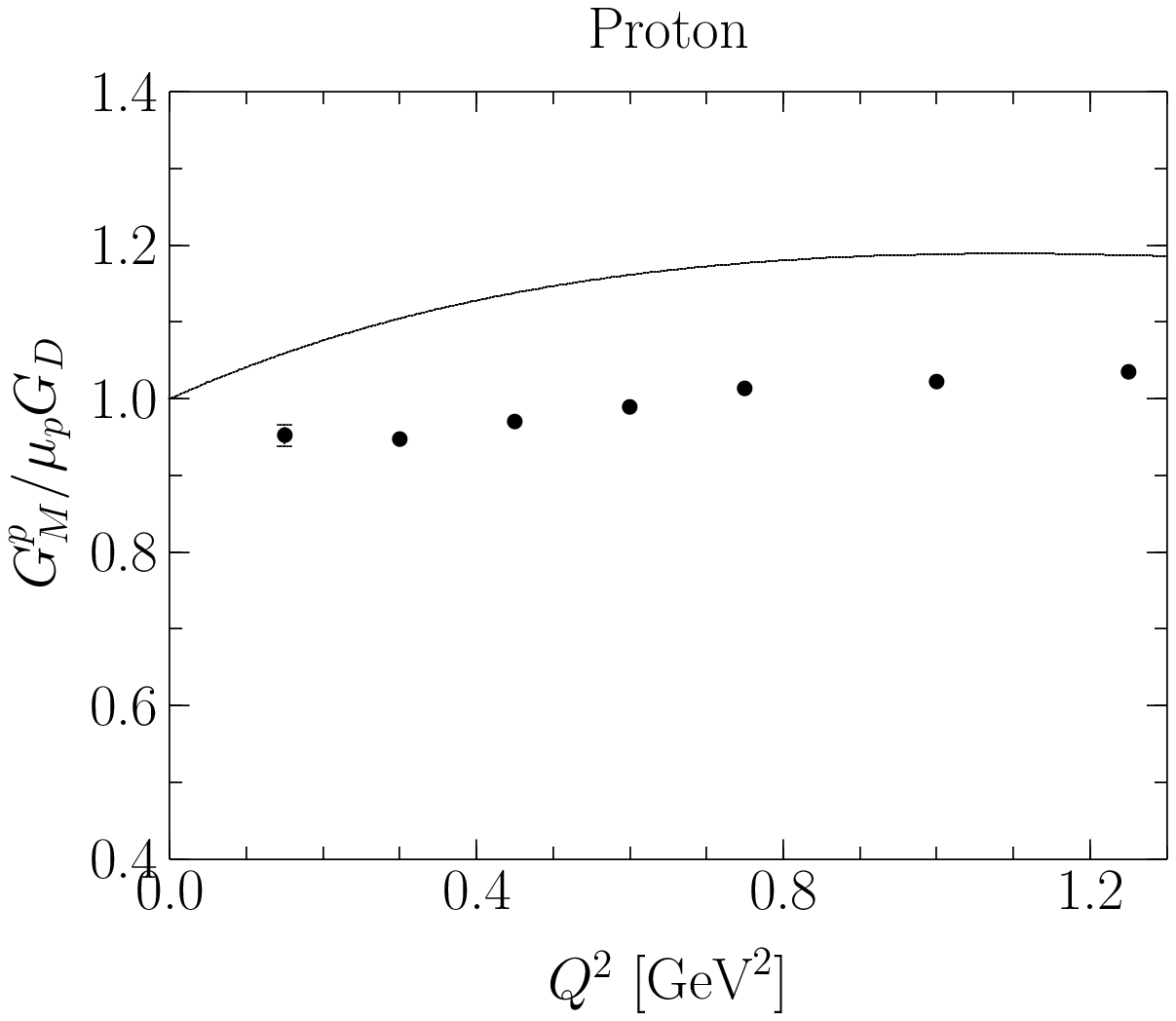}
\hskip -0.00in \includegraphics[scale=0.7]{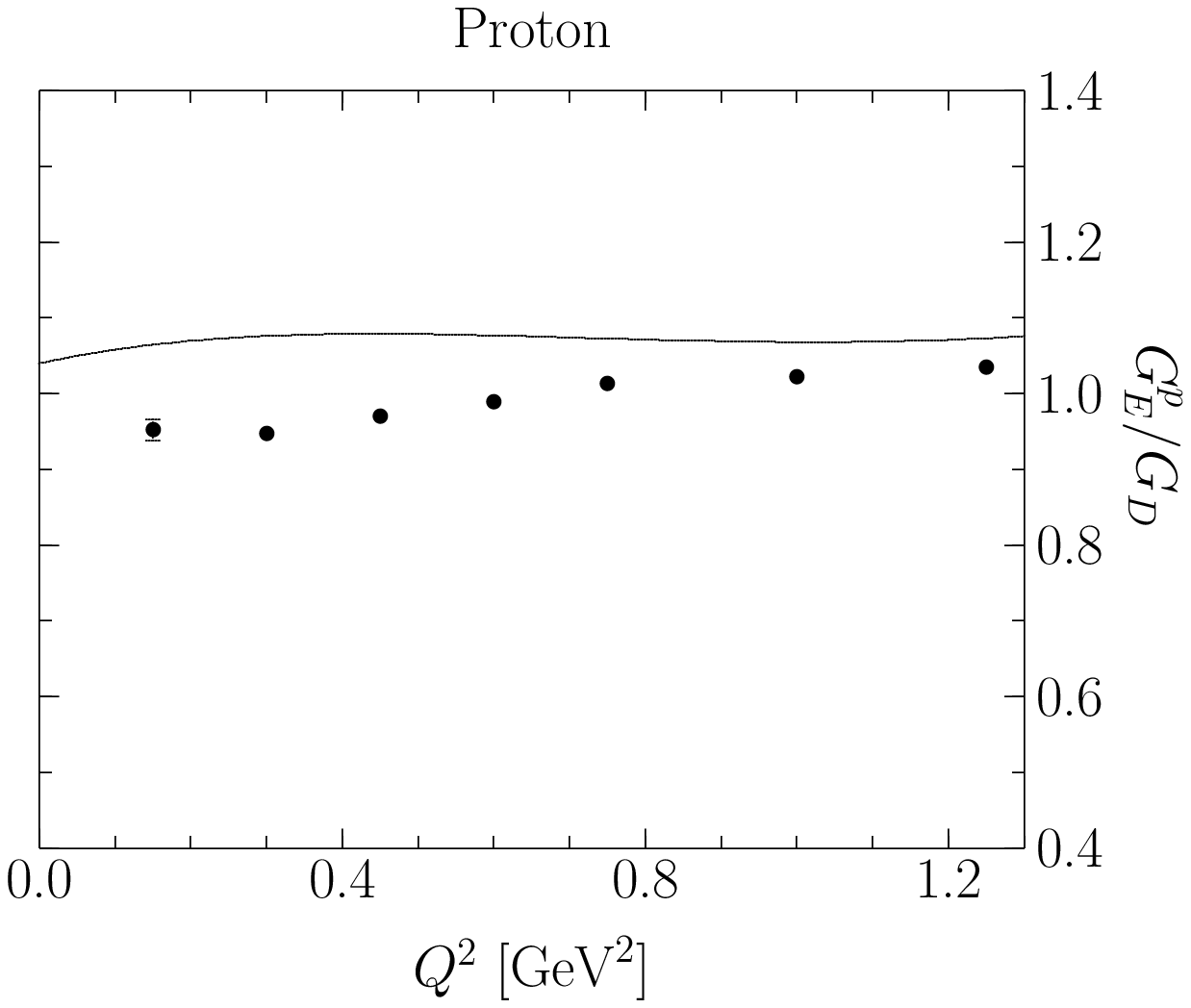}}%
%\hskip -0.00in \includegraphics[scale=0.7]{de_pic.ps}}%
\caption{The predictions are given by the solid lines and the filled
circles are the data taken from~\cite{Walker:1993vj}. The left panel corresponds
to the magnetic form factor of proton and the right panel corresponds to the electric
form factor of proton.}
\label{fig1}
\end{figure}

In the past before the advent of precision data, more recently from the JLab, the standard parametrization of the nucleon form factors was done in terms of a dipole form. The dipole form clearly showed that the nucleon form factors could not be understood in terms of the monopole form alone given by the vector dominance with $V^{gs}$. New data, particularly those from polarization-transfer~\cite{data}, show that the dipole form factor starts deviating for $Q^2 \gsim$ 2 GeV$^2$, so there is nothing sacred with the dipole form. However for $Q^2 \lsim$ 1 GeV$^2$, it is still a good standard measure. We therefore examine the ratios
\be
R^m_2(Q^2)&=&G^p_M (Q^2)/\mu_p G_D (Q^2),\\
R^e_2(Q^2)&=&G^p_E (Q^2)/G_D (Q^2)
\ee
with the dipole form factor parameterized as $G_D=1/(1+Q^2/0.71)^2$. These are plotted in Fig.~\ref{fig1}.
%
%\begin{figure}[ht]
%\vskip 0.1in
%\centerline{\includegraphics[scale=0.6]{dipole_pic.eps}}%
%\caption{The prediction is given by the solid line and the filled
%circles are the data taken from~\cite{Walker:1993vj}.}
% \label{fig1}
%\end{figure}
%
Within the range of momentum transfers considered, the theoretical
$G_M/\mu_p$ is seen to overshoot the experimental by about $\sim
14\%$ and the corresponding $G_E$ by $\lsim 10\%$. These results can
be simply understood by the differences between the predicted radii
$\sim 0.74$ fm and $\sim 0.80$ fm for $G_M$ and $G_E$, respectively
and the radius given by the dipole form factor $\sim$ 0.81 fm. The
conclusion then is that the mechanism that accounts for the defects
in size ${\Delta r}_M\sim 0.07$ fm and ${\Delta r}_E\sim 0.01$ fm are
responsible for the small observed discrepancies.
\subsection{$Q^2 F_2(Q^2)/F_1(Q^2)$}
\begin{figure}[ht]
\centerline{\includegraphics[scale=0.6]{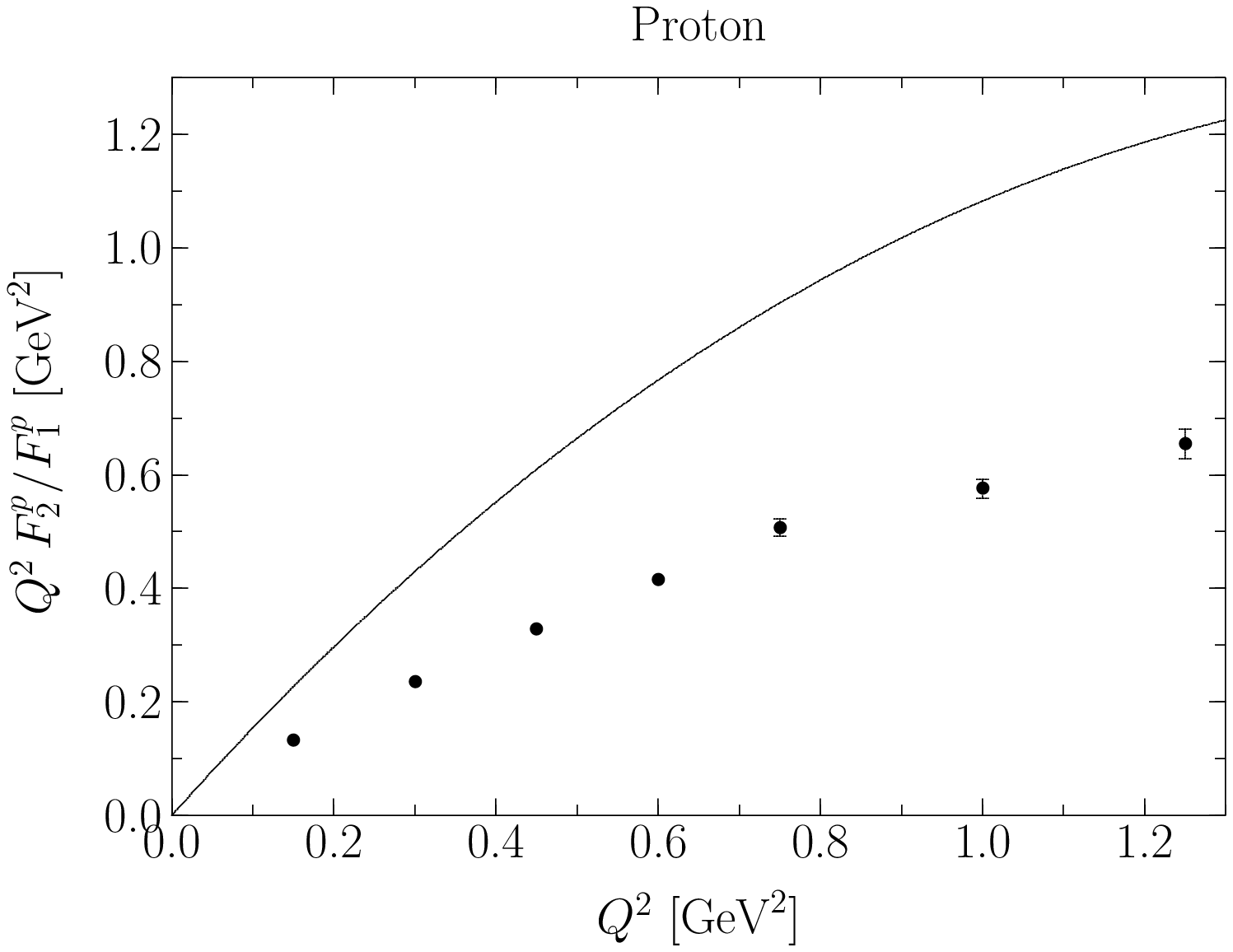}}%
\caption{The  filled circles are the data taken
from~\cite{Walker:1993vj}. The solid line is the prediction in the
SS model with $M_{KK}=0.94$ GeV.  }
 \label{fig3}
\end{figure}

To identify the source of the small deviation observed above, it is instructive to look at the ratio of the Pauli form factor over the Dirac form factor:
\be
R_3 (Q^2)=Q^2 F_2(Q^2)/F_1(Q^2)
\ee
with $Q^2$ given in units of GeV$^2$. This ratio is plotted in Fig.~\ref{fig3}.
We observe that the predicted Pauli form factor drops too slowly
compared with the Dirac form factor at large $Q^2$. As noted above, QCD proper
demands that asymptotically $F_2/F_1\sim Q^{-2}$ but this feature is
missing in hQCD in the large $N_c$ and $\lambda$ limit. We believe
that this feature gives a simple explanation as to how the small
discrepancies arise in Fig.~\ref{fig1}: A reduction of $F_2$ by the
amount indicated in Fig.~\ref{fig3} would bring the theoretical
curves closer to the dipole form factor. Furthermore the
overestimate of $F_2$ for non-zero $Q^2$ can account for the $\lsim
10\%$ undershooting of the ratio $R_1$ as one can see from how $F_2$
enters into $G^p_M$ [Eq.~(\ref{GM})] and $G^p_E$ [Eq.~(\ref{GE})].
\section{Violation of \VMD0}
That hQCD \`a la SS restores vector dominance to the nucleon
structure on the same footing as for the pion raises an interesting
question on the possible role of the higher members of the tower of vector mesons and their relation to the predicted violation~\cite{HS:VD,HS:2007} of \VMD0 in hot and dense matter. This is a current topical
issue in holographic QCD~\cite{Kim:2007xi} prompted by experimental developments on heavy-ion collisions at
CERN and RHIC where hadronic matter is heated to high temperature and compressed to high density.

To bring out the basic issue, it is illuminating to consider the
hidden local symmetry (HLS) theory of Harada and
Yamawaki~\cite{HY:PR} which in some sense can be
interpreted~\cite{HLS-AdS} to be a truncated version of the SS model
in which all higher members of the tower than the ground state vector mesons ($V^{gs}$) are
integrated out and the high-energy sector is matched at a scale
$\sim M_{KK}$ to QCD via various current-current correlators. One
can also view the $V^{gs}$ that figure in HLS theory as ``emergent"
fields~\cite{Kyoto} which when extended to an infinite number, leads
to a dimensionally deconstructed 5D Yang-Mills theory of
QCD~\cite{son-stephanov}. The emergence of local (nonabelian) gauge
invariance is analogous to the emergence of a $U(1)$ gauge degree of
freedom in the CP$^{n-1}$ model. Here the gauge degrees of freedom
are lodged in the chiral field
$U=e^{2i\pi/F_\pi}=\xi_L^\dagger\xi_R$ with $\xi_{L.R}=e^{\mp
i\pi/F_\pi}e^{i\sigma/F_\sigma}$ where $\pi$ is the pion field and
$\sigma$ is the scalar multiplet that makes up the redundant degree
of freedom. This redundancy is elevated to a gauge degree of freedom
with the vectors $V^{gs}$ emerging as a (hidden) gauge field. What
is important for our discussion here is that this HLS theory has a
fixed point to which the system is driven constrained by chiral
symmetry of QCD. At that fixed point (called ``vector manifestation
(VM) fixed point"), the parameters of the HLS Lagrangian take the
values
 \be
(g^*, f_\pi^*, a^*)=(0, 0, 1)\label{vmfp}
 \ee
where $g$ is the hidden gauge coupling, $f_\pi$ is the physical pion
decay constant and $a$ is the ratio of decay constants
 \be
a=\left(\frac{F_\pi}{F_\sigma}\right)^2.
 \ee
The Harada-Yamawaki theory with the VM fixed point (\ref{vmfp}) is
called ``HLS/VM."

In this theory, the isovector photon coupling is given by
 \be
\delta {\cal L}=e{\cal A}_{EM}^\mu\left(-2aF_\pi^2  {\rm
Tr}[g\rho_\mu \hat{Q}] +2i(1-a/2) {\rm Tr}[J_\mu
\hat{Q}]\right)\:,\label{coupling}
 \ee
where $\hat{Q}$ is the quark charge matrix, $\rho_\mu$ is the lowest-lying
iso-vector vector meson and $J_\mu$ is the iso-vector vector current
made up of the chiral field $\xi$. The first term of
(\ref{coupling}) represents the photon coupling through a $\rho$ and
the second term the direct coupling. The vector dominance in this
theory (\VMD0) is obtained when $a=2$ for which the well-known KSRF
relation for the $\rho$ meson holds, i.e., $m_\rho^2=af_\pi^2
g^2=2f_\pi^2 g^2$.

The crucial observation in this theory~\cite{HY:VD} is that $a=2$
that leads to \VMD0 is $not$ on the RG trajectory connected to the
fixed point. In fact, $a=2$ is found to lie on an unstable
trajectory and any infinitesimal perturbation moves the system away
from the value $a=2$. In nature, this is the case when hadronic
matter is heated or compressed~\cite{HS:VD}, with the chiral phase
transition that occurs when the quark condensate goes to zero
coinciding with the $a=1$ point. At this point, \VMD0 is maximally violated.

We can now interpret what we found in the previous section in hQCD
in terms of HLS/VM as follows based on two observations: (a) In
matter-free space, the pion form factor is given by $a=2$, so it is
vector-dominated in the Sakurai sense. What is somewhat surprising
is that the \VMDinf in hQCD (nearly completely saturated by the four lowest members of
the tower) also describes the pion form factor well. Here the two
pictures seem to give quite similar results for the pion form
factors. However this may be coincidental, for the RG analysis has
shown that when the system is heated and/or compressed, $a$ departs
quickly from 2 and moves to the fixed point $a=1$~\cite{HY:PR}
thereby violating \VMD0; (b) as for the nucleon, the \VMD0 is maximally violated already in matter-free vacuum: the two-component models imply that $a\approx 1$, with a
phenomenologically favored 50-50 coupling to the vector meson and
the ``core" represented by the second term of (\ref{coupling}) as
discussed in \cite{chiralbag}.

The above two observations are suggestive of that the higher members
of the tower in hQCD could be playing an important role in hot/dense
medium. The key observation in HLS/VM is the decrease of
the quark condensate $\la\bar{q}q\ra$ as well as of the vector meson
mass $m_{V^{gs}}$ as temperature/density increases. But this feature is missing in the current work on hQCD of SS in medium in which the large $N_c$ and $\lambda$ limit is taken and the current quark mass is left out. It would be interesting to see how the missing ingredient in the core size we find in the nucleon form factor is related to the chiral properties of hadrons seen in Harada-Yamawaki's HLS/VM theory.
\section{Further Comments}
In this paper, we made a simple evaluation of the nucleon form
factors that are vector-dominated by the infinite tower of vector
mesons as derived from the instanton solution in the SS
model~\cite{Hong:2007ay}. With the infinite tower truncated to the
four lowest vector mesons that saturate within a few \%
the zero-momentum sum rules, the (parameter-free) results, e.g., the
proton radii, the proton form factors to the momentum transfer
restricted by the KK mass $M_{KK}\sim 1$ GeV etc., come out to fare
well with experiments. We could have done much better in comparing
with the data by implementing ad-hoc phenomenological form factors
that simulate the asymptotic freedom structure of QCD as has been
done in the two-component models~\cite{IJL} and in the Skyrme
model~\cite{holzwarth}. That would have allowed us to go beyond the
kinematic regime $Q^2\lsim 1$ GeV$^2$ and get a much better fit. But
this was not the aim of our work. Our aim was to see whether the
hQCD model as defined in given approximations and free of unknown
parameters can resemble Nature. Our results answer this
question in the positive and indicate how to improve the comparison
with nature.

Assuming that the model $can$ make meaningful predictions in the
regime that QCD proper is unable to access, an interesting question
to ask is what issues can be profitably addressed by the hQCD model.
Indeed one of such issues is the role of the infinite tower in hot
and dense medium discussed in the preceding section. Specifically it
would be exciting if one could study how hadrons behave as
temperature/density approaches the critical point where chiral phase
transition is presumed to take place, currently a hot topic in the
AdS/QCD circle~\cite{Horigome:2006xu}. In HLS/VM theory of Harada and Yamawaki, the strong
violation of vector dominance with \VMD0 near the critical point is
closely linked to the properties of the hadrons involved, e.g.,
vanishing vector meson ($\rho$, $\omega$) masses and pion decay
constant etc. Whether or not this description by HLS/VM theory is
correct cannot at the moment be assessed by QCD: There are no QCD
tools, including lattice, that allow one to access that regime. It
seems plausible that in hQCD, it is the tower that will replace the
role of $a$ in HLS theory. Since in HLS/VM theory it is the quark
condensate $\la\bar{q}q\ra$ that plays the key role, one would have
to figure out how to correctly introduce the quark masses and quark
condensates in hQCD to address the issue.

\subsection*{Acknowledgments}
M.R. is grateful for discussions with Masa Harada and
Koichi Yamawaki on vector dominance. D.K.H. and P.Y. thank Yukawa Institute
of Theoretical Physics for the hospitality and acknowledge illuminating
conversations with Kenji Fukushima, Teiji Kunihiro, Chong-Sa Lim and Shigeki Sugimoto.
This work is supported in part (D.~K.~H. and P.~Y.)
by the Korea Research Foundation Grant funded by the
Korean Government (MOEHRD, Basic Research Promotion Fund)
(KRF-2007-314-C00052),
(D.~K.~H) by KOSEF
Basic Research Program with the grant No. R01-2006-000-10912-0, (P.~Y) by
the Science Research Center Program of
the Korea Science and Engineering Foundation through
the Center for Quantum Spacetime(CQUeST) of
Sogang University with grant number R11-2005-021.
H.U.Y. is partly supported by the Korea Research Foundation Grant KRF-2005-070-c00030,
and thanks Ki-Myeong Lee for a financial support from his fund.

%\vfill
%\eject

\end{document}